\documentclass{article}

\usepackage[final]{my_nips_2016}

\usepackage[utf8]{inputenc} 
\usepackage[T1]{fontenc}    
\usepackage{hyperref}       
\usepackage{url}            
\usepackage{booktabs}       
\usepackage{amsfonts}       
\usepackage{nicefrac}       
\usepackage{microtype}      
\usepackage{graphicx}       
\usepackage{tikz}

\title{Characterizing Allegheny County Opioid Overdoses with an Interactive Data Explorer and Synthetic Prediction Tool}

\author{
  Theresa Gebert\\
  Department of Statistics and Data Science\\
  Carnegie Mellon University\\
  \texttt{theresa@stat.cmu.edu} \\
  \And
  Shuli Jiang \\
  Department of Computer Science\\
  Carnegie Mellon University\\
  \texttt{shulij@andrew.cmu.edu} \\
  \AND
  Jiaxian Sheng \\
  Department of Computer Science\\
  Carnegie Mellon University\\
  \texttt{jiaxians@andrew.cmu.edu} \\
}

\begin{document}

\maketitle

\begin{abstract}
The United States has an opioid epidemic, and Pennsylvania's Allegheny County is among the worst. This motivates a deeper exploration of what characterizes the epidemic, such as what are risk factors for people who ultimately overdose and die due to opioids. We show that some interesting trends and factors can be identified from openly available autopsy data, and demonstrate the power of building an interactive data exploration tool for policy makers. However, there is still a pressing need to incorporate further demographic factors. We show this by using synthetic Electronic Medical Record (EMR) data to simulate the predictive power of random forests and neural networks when given additional loosely correlated features. In addition, we give examples of useful feature extraction that enable model enhancement without sacrificing privacy.
\end{abstract}

\section{Introduction}
There is an opioid epidemic in the United States \citep{alexander15}, motivating extensive analysis in the last two years alone \citep{sankaran16, chen17, neill17, battista17}. In order to best serve their communities, policy-makers need to know a variety of things about this epidemic, such as:
\begin{enumerate}
\item What exactly is happening right now: who is affected and how much?
\item Why and how did this happen: what factors influenced increased opioid usage?
\item What can be done to slow and stop this epidemic?
\item How can we prevent this from happening again in the future?
\end{enumerate}
Data and statistical analysis should be able to help answer some of these questions, and machine learning may be able to create tools to aid prevention efforts. In this paper, we showcase the work we did during the 2-day \href{https://www.hackauton.com/}{2018 HackAuton} hackathon. Policy-makers frequently do not have the statistics or computer science background required to easily manipulate and explore data, or easily create and interpret models. That is why our first goal was to create an interactive data explorer, that provides a simple, intuitive user interface for the key stakeholders, and a lightweight, cheap backend in \texttt{RShiny} to enable easy codebase maintenance. Our second goal was to create a synthetic prediction tool: something that would show policy-makers the power of additional data and allow them to easily communicate this to others. In the last section of the paper, we also describe our desired future work and policy recommendations.

\section{Background}
Opioids are powerful painkillers, which have traditionally made them an attractive drug to treat patients with severe pain or to manage pain after surgery. Well-known opioids include heroin and morphine. When opioid drugs bind to opioid receptors in the brain, they can drive up dopamine levels and produce a state of euphoria and relaxation \citep{nids16}. While the effects include euphoria, they also include drowsiness, nausea, confusion, constipation, sedation, tolerance, addiction, respiratory depression and arrest, unconsciousness, coma, and death. Opioids are highly addictive substances, which is why doctors recommend they only be used in the extreme cases. However, the number of prescriptions of opioids to take at home has also risen over the last decade, which makes it plausible that increased rates of prescription opioids is leading to initial addiction and driving the rates of illegal opioid usage. Some researchers also link stringent insurance policies to the epidemic \citep{times17}. Opioids costs the United States billions of dollars every year in direct healthcare costs and lost productivity. Some of the states that are most affected are Massachusetts, Connecticut, and Pennsylvania \citep{hedegaard17}.

The \href{https://www.alleghenycounty.us/medical-examiner/index.aspx}{Allegheny County Medical Examiner's Office} has made autopsy data from fatal accidental overdose cases \href{https://catalog.data.gov/dataset/allegheny-county-fatal-accidental-overdoses}{publicly available} from the years 2008 thru 2017. Each fatal overdose incident is in Allegheny County and includes the age, sex\footnote{Sex data was recorded on a binary scale, which does not include individuals with uncertain or non-binary sex. While the paper may reference ``both sexes,'' the authors intend this as a statement about the sex measurements included in the dataset, and not all possible sexes in the human population. In addition, the authors recognize that sex does not capture gender identity.}, race, drugs present\footnote{In particular, up to seven drugs identified as cause-of-death related to the overdose.}, zip code of incident\footnote{Note that zip code of incident is where the Office of the Medical Examiner received the body, not necessarily where the overdose occurred.} and zip code of residence. The dataset also contains the date and time of death. It only includes cases considered closed and some features are missing or incorrect (e.g. 4-digit zip code). There are a total of $3483$ cases.

\section{Interactive Data Explorer}
For all but the most simple datasets, the vast majority of time in model construction and analysis is spend on data exploration and visualization. Simply subsetting the data and extracting features from raw input can require formal computer science training. Not all offices have an in-house statistician or programmer; even if they do, these analysts often have many responsibilities; and hiring an external consultant is time-consuming and expensive. Giving policy-makers the ability to easily interact with data, ask and answer questions, and communicate key insights is crucial. As a result, we had three major goals in constructing our interactive data exploration tool:
\begin{enumerate}
\item interactive;
\item easy and intuitive to use;
\item lightweight and cheap to maintain.
\end{enumerate}

\subsection{Methods}
Given our own time constraints, we had a useful test case for satisfying our third engineering goal: lightweight and cheap to maintain. Our hypothesis was that if we can build this tool in a day, it will be easy for a high school intern at the Allegheny County Medical Examiner's Office to maintain, or an overworked in-house IT employee. We also used ourselves as guinea pigs in learning the tool, since none of us have experience with front-end development and design. Our tool of choice ended up being \texttt{RShiny}.

\begin{figure}[h]
\centering
\includegraphics[width=0.75\textwidth]{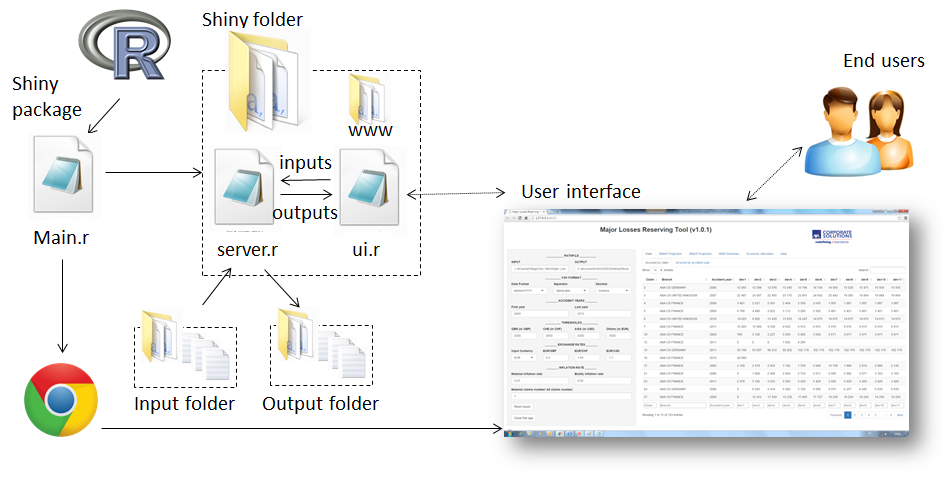}
\caption{\texttt{RShiny} is a lightweight tool for building beautiful interactive web applications to visualize data. \texttt{R} calls the Shiny app in a main script with the Shiny package. Image courtesy of Little Actuary.}
\label{fig:shiny_structure}
\end{figure}

In less than one hour, we had \texttt{R} and the requisite packages installed and a working version of the \texttt{RShiny} tutorial. It only took us another hour to build our first interactive data visualization using the Allegheny County Overdose dataset. This highlights the usability of the product and its promise as a feasible, maintainable tool within a government organization ~\ref{fig:shiny_structure}.

\subsection{Results}

\begin{figure}[h]
\centering
\includegraphics[width=0.5\textwidth]{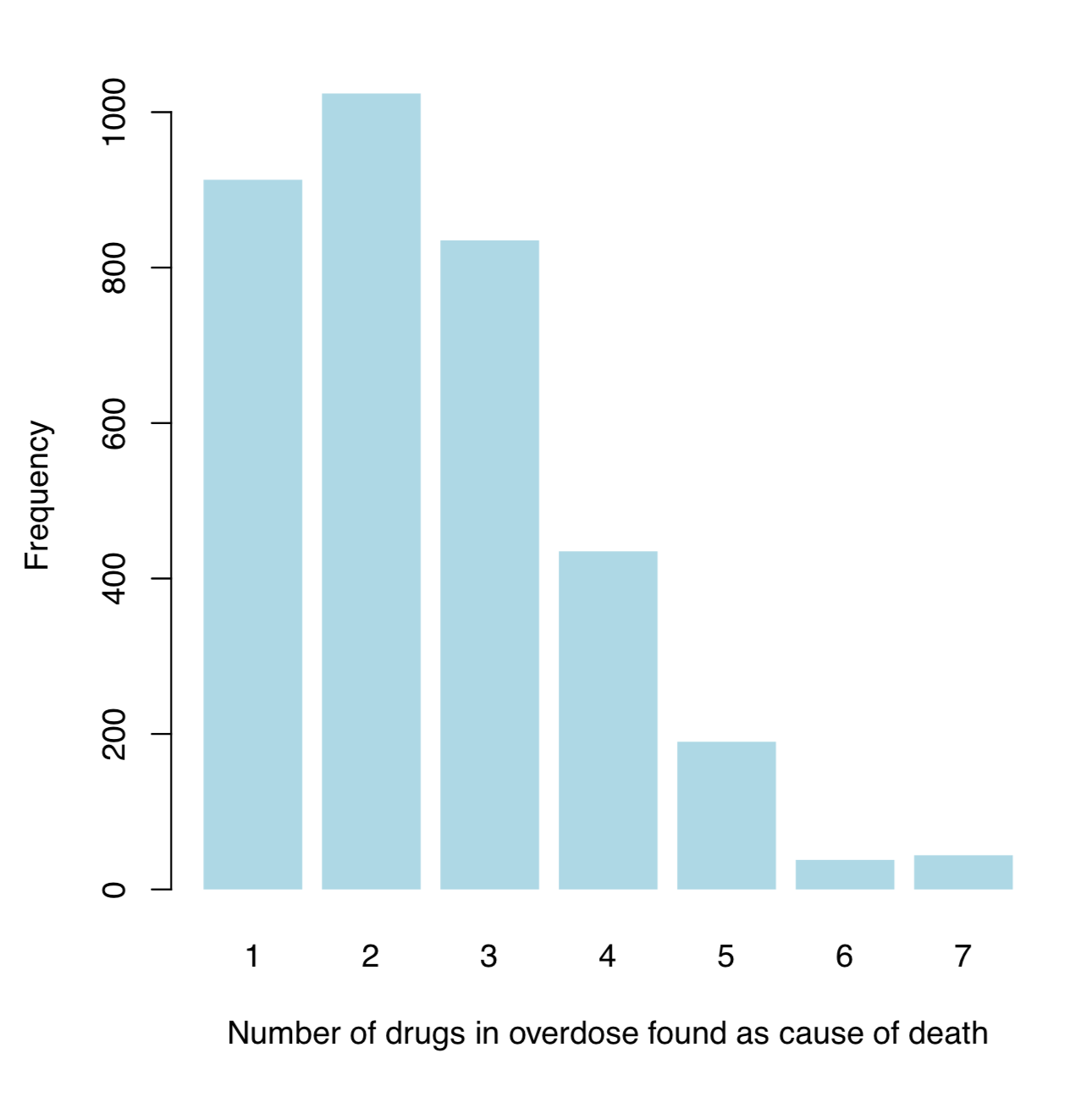}
\caption{The frequencies of the number of drugs involved in an overdose aggregated across all cases and all years. Having at least 2 drugs involved in an overdose is extremely common.}
\label{fig:numdrugs}
\end{figure}

\begin{figure}[h]
\centering
\includegraphics[width=0.75\textwidth]{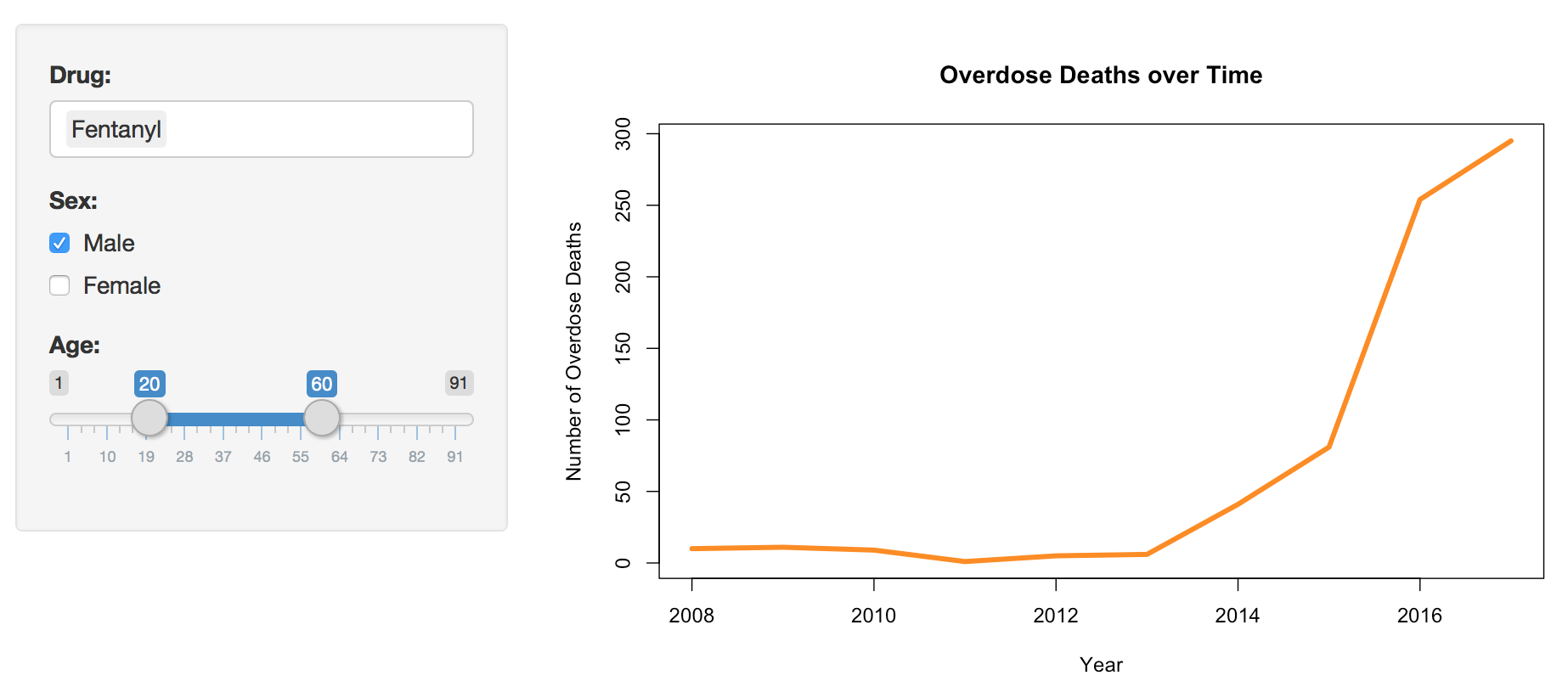}
\caption{The first interactive graph we constructed for the tool. It enables the visualization of overdose deaths over time constrained to different variables: by drug, sex, and age. Drug and sex allow multiple selections. Age allows selection of an age range with variable length.}
\label{fig:timeline}
\end{figure}

Our data exploration tool provides a few static graphs to highlight some known features of interest, such as top drugs used across all cases for all years, and geographic distribution. Our interactive graphs enable exploring more complex relationships, like change over time, and co-occurrence between drugs. From our exploratory analysis we discovered some simple characteristics:
\begin{itemize}
\item {\bf opioid-related overdoses have been significantly increasing over time}; this is as expected, since this is what motivates this entire project ~\ref{fig:timeline}.
\item {\bf overdoses are frequently related to many drugs}; the mean number of drugs involved in an overdose across all cases, all years, for both sexes, was 2.5. A density plot of the number of drugs involved shows that 2 drugs are most frequently involved ~\ref{fig:numdrugs}.
\item {\bf a few drugs are involved in most of the cases}; this is what we might expect, but the data confirms that the top 8 drugs account for over 75\% of cases.
\item {\bf cases are geographically concentrated in more populous area}; again, this is what we might expect, but what is more interesting is actually that, relative to population, the frequency of overdoses in rural and less populated areas is higher than in urban areas.
\end{itemize}

\begin{figure}[h]
\centering
\includegraphics[width=0.75\textwidth]{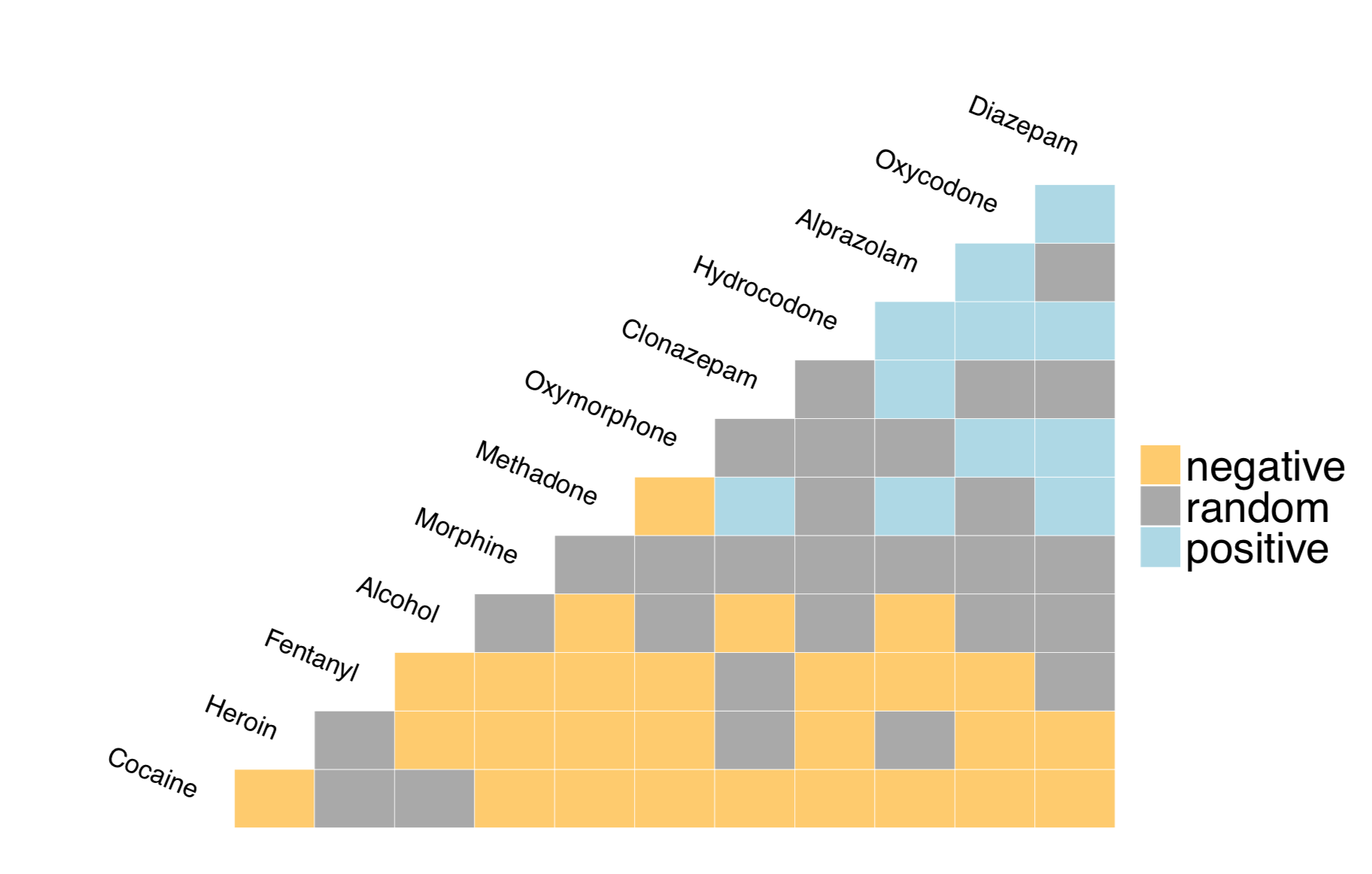}
\caption{The cooccurrence of pairwise drugs for the top drugs used. Blue indicates those two drugs are frequently found together in an overdose, yellow indicates those two drugs tend not to be found together in an overdose.}
\label{fig:corr}
\end{figure}

From our exploratory analysis we also discovered some more nuanced phenomena:
\begin{itemize}
\item {\bf Fentanyl is a major player in opioid use increases}; Fentanyl is a powerful synthetic opioid analgesic that is similar to morphine but is 50 to 100 times more potent \citep{nids16}. Its usage saw a major spike in 2015 across all races, age groups, and both sexes. Contrast this with acetaminophen, which has been in popular use for decades and usage associated with overdose is both overall low, and has not increased.
\item {\bf hard drugs tend not to be taken together}; while we may expect this behavior given anecdotal evidence and economic viability (illicit drugs are expensive), the data shows a clear trend here: the drugs that appear most frequently in the overdoses are also hard drugs (e.g. heroin, cocaine), and they are less likely to be taken together ~\ref{fig:corr}.
\item {\bf opioids, stimulants, and depressants tend not to be taken together}; again, while we may expect this behavior given anecdotal evidence, the data shows a clear trend here: classifying drugs within their major categories (opioids, stimulants, depressants) reveals that these drugs are more likely to be mixed within-category than between-category.
\end{itemize}
The cooccurence visualization was produced using the \texttt{cooccur} package in \texttt{R} \citep{griffith16}. The package is based on probabilistic models developed in ecology to test for statistically significant pair-wise patterns of species cooccurrence in different locations \citep{veech12}. This particular model has the advantages of being metric-free, distribution-free, and randomization-free. The model gives the probability that two drugs would cooccur at a frequency less than (or greater than) the observed frequency if the two drugs were distributed independently of one another. The null hypothesis is that pairs of drugs are taken together uniformly randomly.

The way that these results can inform policy decisions is discussed in the last section of this paper.

\section{Synthetic Prediction Tool}
The security and confidentiality of medical data is of high importance. This is why everything from conversational discretion by healthcare providers and the security of medical records is highly regulated and strictly enforced. While patient privacy is becoming increasingly important, especially with regards to disclosure to insurance companies and employers, it can also slow and hamper progress in other ways. For example, the only patient information we have about the Allegheny County overdose cases is age, sex, and race. While these factors can be important factors in overdose risk, other factors may be far better predictors and improve our ability to identify individuals at risk of drug use, and at risk of dying from overdose. However, in order for such prediction models to be useful to social workers, policy-makers, and the patients themselves, they need to be easy to use and dynamically adapt to different patient characteristics. Finally, a major pitfall of traditional risk analysis tools is that they only provide point estimates, without giving any measure of uncertainty in their estimate. Therefore, we had three major goals in constructing our synthetic prediction tool:
\begin{enumerate}
\item dynamic;
\item easy and intuitive to use;
\item incorporates a measure of uncertainty.
\end{enumerate}

\subsection{Methods}
In order to build our prediction tool, we needed to (1) generate synthetic data for a non-overdose population, (2) generate additional medical predictors, and (3) build a predictive model. Let $\mathcal{X}_o \in \mathbb{R}^{n_o,d}$ denote the $n_o$ observations of $d$-dimensional features of the population that died from overdose. Let $\mathcal{X}_c \in \mathbb{R}^{n_c,d}$ denote the $n_c$ observations of $d$-dimensional features of the population that does not die from overdose. Notice that our control population could be one of two types, both interesting: people in the general population, both users of opioids and non-users of opioids, or people in the opioid-using population. For simplicity, we generated synthetic data for the general population, but it would be very useful to know the risk factors for an individual to overdose given that they are already using drugs.

Our available dataset is $\mathcal{X}_o$, so we do not need to generate that. Our first challenge is generating $\mathcal{X}_c \in \mathbb{R}^{n_c,d}$. We took advantage of a free synthetic Electronic Medical Record (EMR) database online in order to obtain the demographic information of the general population, under the assumption that the simulated data was representative of the general population. \href{https://synthetichealth.github.io/synthea/#home}{Synthea} is an open-source, synthetic patient generator that models the medical history of synthetic patients \citep{walonoski18}. For each of 1000 synthetic patient, Synthea data contains a complete medical history, including medications, allergies, medical encounters, and social determinants of health. This data can be used without concern for legal or privacy restrictions. Other options for synthetic data we considered included another source of synthetic medical data \citep{kartoun16} and a source of synthetic population data that would enable county-level specificity \citep{gallagher17}.

Our second challenge was to extract additional features, to show the prediction improvement in our model in the presence of additional covariates. In more precise terms, we transformed $\mathcal{X}_o \in \mathbb{R}^{n_o,d} \to \mathcal{X}_o \in \mathbb{R}^{n_o,d'}$, where $d' > d$. Here are some of the features we extracted from the medical records:
\begin{itemize}
\item {\bf marital status}; whether an individual is single, married, divorced, or widowed.
\item {\bf socioeconomic status}; percentage of income below poverty line.
\item {\bf language}; primary language spoken at home (English, Icelandic, or Spanish).
\item {\bf sickliness}; time-discounted days spent in hospital.
\item {\bf disease history}; the number of occurrences in each of 20 hand-selected categories of disease as defined by \href{http://apps.who.int/classifications/icd10/browse/2016/en}{ICD-10}.
\end{itemize}

An additional challenge in generating these features was also {\it matching} them to the appropriate overdose cases. Given our time constraints, we used a very simple matching algorithm and have ideas for improvement described in the last section of the paper. The current implementation of our matching procedure works as follows: for each individual that overdosed, in a random order, select a uniformly random individual from the synthetic patients that matches
\begin{enumerate}
\item exactly on gender; this seemed important to match on exactly, but resulted in dropping individuals with missing gender.
\item if race is white, black, or Asian, then match exactly, otherwise match randomly; specifically, for individuals marked as \texttt{Other} or \texttt{Unidentified} we randomly selected a race from the synthetic population.
\item within 3 years on age; this was more difficult to match on exactly, and $\pm$3 years seemed a reasonable first step.
\item and then remove that selected individual from the possible candidates; this is done to prevent duplicate records.
\end{enumerate}

Finally, the last step was building our models. We built a simple model to predict the number of drugs used in the overdose using only the original data. Then, we build predictive models using the $n_0$ overdose cases with the $d$ original feature set and with the $d' > d$ enhanced feature set to showcase the improvement of additional information.

\subsection{Results}

\begin{figure}[h]
\centering
\includegraphics[width=0.5\textwidth]{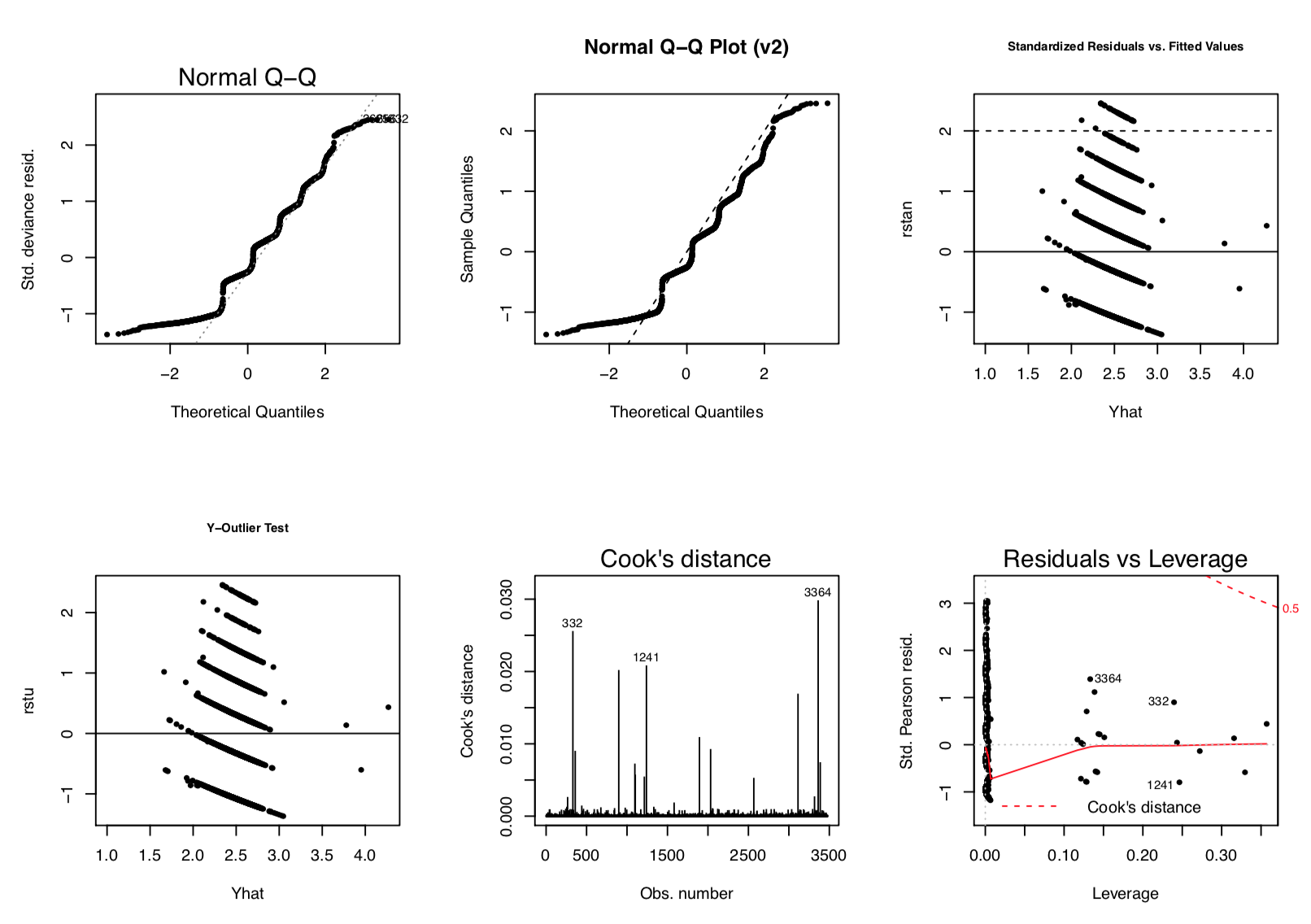}
\caption{Diagnostic plots for the Poisson generalized linear model. The assumption of Normality of errors is not perfect, but reasonable. No apparent outliers are detected. Despite the odd appearance of the residual plots and the slight positive skew, this is actually a reasonable fit.}
\label{fig:diags}
\end{figure}

Our first model was a Poisson generalized linear model with the canonical link function to associate the outcome, number of drugs involved in overdose, with three features of interest: age, sex, and race. We did not include interaction terms due to the small size of our dataset relative to the number of features (sex and race are categorical variables and generate several indicator variables in the model). We did not run additional models thereby avoiding multiple comparisons problems. Our model summary is shown in ~\ref{table:glmSum}. We performed a Goodness of Fit test and reject the null hypothesis that the model does not explain the data well. The diagnostic residual and QQ plots also looked reasonable ~\ref{fig:diags}. We also performed a Deviance Test to assess overdispersion and concluded that the Poisson model fit the data well and there was no overdispersion.

\begin{table}[ht]
\centering
\begin{tabular}{rrrrr}
  \hline
 & Estimate & Std. Error & z value & Pr($>$$|$z$|$) \\ 
  \hline
(Intercept) & 0.6871 & 0.0548 & 12.54 & {\bf 0.0000} \\ 
  as.factor(Sex)Male & -0.0353 & 0.0230 & -1.54 & 0.1243 \\ 
  as.factor(Race)White & 0.1390 & 0.0329 & 4.23 & {\bf 0.0000} \\ 
  as.factor(Race)Hispanic & -0.0946 & 0.2519 & -0.38 & 0.7072 \\ 
  as.factor(Race)Other & -0.0823 & 0.3550 & -0.23 & 0.8167 \\ 
  as.factor(Race)Asian & -0.2169 & 0.2907 & -0.75 & 0.4555 \\ 
  as.factor(Race)Middle Eastern & 0.5753 & 0.2902 & 1.98 & {\bf 0.0475} \\ 
  as.factor(Race)Unidentified & -0.8342 & 1.0005 & -0.83 & 0.4044 \\ 
  as.factor(Race)Indian & -0.0258 & 0.7082 & -0.04 & 0.9709 \\ 
  Age & 0.0032 & 0.0009 & 3.66 & {\bf 0.0003} \\ 
   \hline
\end{tabular}
\label{table:glmSum}
\end{table}

As we can see from the coefficients, age is slightly but significantly positively associated with number of drugs used, even when accounting for the other variables. In addition, being white is slightly but significantly positively associated the number of drugs used in the overdose.

Finally, we used the synthetic data to develop a predictive model of risk of overdose with and without our new features. As expected, the new features improved the model accuracy. In particular, we see that given our original, minimal covariates, model accuracy is little better than random chance (baseline is 50\% by construction). The outcome of each patient is whether or not he or she dies from drug overdose. Meanwhile, with the features from the EMR data, we see a mean accuracy of the model for this binary classification increases to 92\% for neural network and 97\% for random forest. We are still investigating the discrepancies between these models.

We fit a random forest on these features and performed binary classification on whether or not an individual would overdose on drugs. The model achieves a sensitivity of 97.2\% and specificity of 97.2\% using 10-fold cross-validation. We found that age, race and the language spoken by the person are the most important features, each contributing 44\%, 19\% and 10\% to the final prediction.

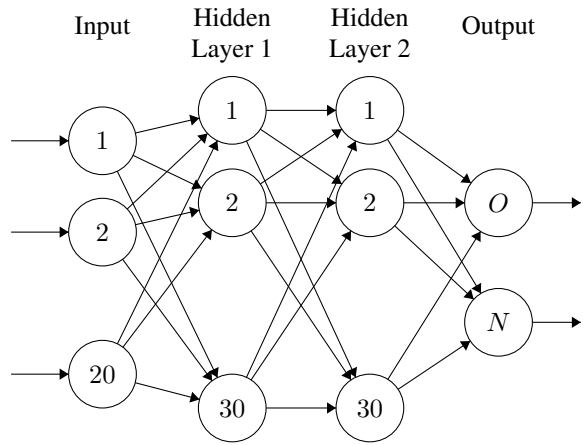
\begin{figure}[h]
\centering
\begin{tikzpicture}[scale=0.15]
\tikzstyle{every node}+=[inner sep=0pt]
\draw [black] (22.6,-21.5) circle (3);
\draw (22.6,-21.5) node {$1$};
\draw [black] (22.6,-29.6) circle (3);
\draw (22.6,-29.6) node {$2$};
\draw [black] (22.6,-42.2) circle (3);
\draw (22.6,-42.2) node {$20$};
\draw [black] (34.1,-18.8) circle (3);
\draw (34.1,-18.8) node {$1$};
\draw [black] (34.1,-27) circle (3);
\draw (34.1,-27) node {$2$};
\draw [black] (34.1,-45.2) circle (3);
\draw (34.1,-45.2) node {$30$};
\draw [black] (46.3,-18.8) circle (3);
\draw (46.3,-18.8) node {$1$};
\draw [black] (46.3,-27) circle (3);
\draw (46.3,-27) node {$2$};
\draw [black] (46.3,-45.2) circle (3);
\draw (46.3,-45.2) node {$30$};
\draw [black] (57.7,-27) circle (3);
\draw (57.7,-27) node {$O$};
\draw [black] (57.7,-37.6) circle (3);
\draw (57.7,-37.6) node {$N$};
\draw (22.6,-11.5) node {Input};
\draw (34.1,-10.5) node {Hidden};
\draw (34.1,-13.5) node {Layer 1};
\draw (46.3,-10.5) node {Hidden};
\draw (46.3,-13.5) node {Layer 2};
\draw (57.7,-11.5) node {Output};
\draw [black] (60.7,-37.6) -- (65,-37.6);
\fill [black] (65,-37.6) -- (64.2,-37.1) -- (64.2,-38.1);
\draw [black] (60.7,-27) -- (65,-27);
\fill [black] (65,-27) -- (64.2,-26.5) -- (64.2,-27.5);
\draw [black] (48.74,-20.55) -- (55.26,-25.25);
\fill [black] (55.26,-25.25) -- (54.91,-24.38) -- (54.32,-25.19);
\draw [black] (49.3,-27) -- (54.7,-27);
\fill [black] (54.7,-27) -- (53.9,-26.5) -- (53.9,-27.5);
\draw [black] (48.8,-43.54) -- (55.2,-39.26);
\fill [black] (55.2,-39.26) -- (54.26,-39.29) -- (54.82,-40.12);
\draw [black] (48.5,-29.04) -- (55.5,-35.56);
\fill [black] (55.5,-35.56) -- (55.26,-34.65) -- (54.58,-35.38);
\draw [black] (47.86,-21.37) -- (56.14,-35.03);
\fill [black] (56.14,-35.03) -- (56.16,-34.09) -- (55.3,-34.61);
\draw [black] (47.89,-42.66) -- (56.11,-29.54);
\fill [black] (56.11,-29.54) -- (55.26,-29.95) -- (56.11,-30.49);
\draw [black] (37.1,-18.8) -- (43.3,-18.8);
\fill [black] (43.3,-18.8) -- (42.5,-18.3) -- (42.5,-19.3);
\draw [black] (36.59,-20.47) -- (43.81,-25.33);
\fill [black] (43.81,-25.33) -- (43.43,-24.47) -- (42.87,-25.3);
\draw [black] (35.36,-21.52) -- (45.04,-42.48);
\fill [black] (45.04,-42.48) -- (45.16,-41.54) -- (44.25,-41.96);
\draw [black] (36.59,-25.33) -- (43.81,-20.47);
\fill [black] (43.81,-20.47) -- (42.87,-20.5) -- (43.43,-21.33);
\draw [black] (37.1,-27) -- (43.3,-27);
\fill [black] (43.3,-27) -- (42.5,-26.5) -- (42.5,-27.5);
\draw [black] (35.77,-29.49) -- (44.63,-42.71);
\fill [black] (44.63,-42.71) -- (44.6,-41.77) -- (43.77,-42.32);
\draw [black] (37.1,-45.2) -- (43.3,-45.2);
\fill [black] (43.3,-45.2) -- (42.5,-44.7) -- (42.5,-45.7);
\draw [black] (35.77,-42.71) -- (44.63,-29.49);
\fill [black] (44.63,-29.49) -- (43.77,-29.88) -- (44.6,-30.43);
\draw [black] (35.36,-42.48) -- (45.04,-21.52);
\fill [black] (45.04,-21.52) -- (44.25,-22.04) -- (45.16,-22.46);
\draw [black] (25.52,-20.81) -- (31.18,-19.49);
\fill [black] (31.18,-19.49) -- (30.29,-19.18) -- (30.51,-20.16);
\draw [black] (25.31,-22.79) -- (31.39,-25.71);
\fill [black] (31.39,-25.71) -- (30.89,-24.91) -- (30.46,-25.81);
\draw [black] (23.91,-24.2) -- (32.79,-42.5);
\fill [black] (32.79,-42.5) -- (32.89,-41.56) -- (31.99,-42);
\draw [black] (24.79,-27.55) -- (31.91,-20.85);
\fill [black] (31.91,-20.85) -- (30.99,-21.04) -- (31.67,-21.77);
\draw [black] (25.53,-28.94) -- (31.17,-27.66);
\fill [black] (31.17,-27.66) -- (30.28,-27.35) -- (30.5,-28.33);
\draw [black] (24.38,-32.01) -- (32.32,-42.79);
\fill [black] (32.32,-42.79) -- (32.25,-41.84) -- (31.44,-42.44);
\draw [black] (23.92,-39.51) -- (32.78,-21.49);
\fill [black] (32.78,-21.49) -- (31.98,-21.99) -- (32.87,-22.43);
\draw [black] (24.41,-39.81) -- (32.29,-29.39);
\fill [black] (32.29,-29.39) -- (31.41,-29.73) -- (32.21,-30.33);
\draw [black] (25.5,-42.96) -- (31.2,-44.44);
\fill [black] (31.2,-44.44) -- (30.55,-43.76) -- (30.3,-44.72);
\draw [black] (14.5,-42.2) -- (19.6,-42.2);
\fill [black] (19.6,-42.2) -- (18.8,-41.7) -- (18.8,-42.7);
\draw [black] (14.5,-29.6) -- (19.6,-29.6);
\fill [black] (19.6,-29.6) -- (18.8,-29.1) -- (18.8,-30.1);
\draw [black] (14.5,-21.5) -- (19.6,-21.5);
\fill [black] (19.6,-21.5) -- (18.8,-21) -- (18.8,-22);
\end{tikzpicture}
\caption{Structure of the neural network model, which we used to predict the binary outcome overdose vs. not overdose, given a set of features.}
\label{fig:NN}
\end{figure}

The neural network model has three hidden layers: the first hidden layer has 20 neurons, with activation function $\mathsf{ReLU}$ ~\ref{fig:NN}. The second hidden layer has 30 neurons with activation function $\mathsf{tanh}$ and the last hidden layer has 30 neurons with activation function $\mathsf{tanh}$. The output layer has a $\mathsf{softmax}$ activation function and outputs 2 neurons indicating the probability the patient overdoses and dies versus the probability that the patient lives. Again, we used 10-fold cross-validation, and this model achieves a sensitivity of 91.7\% and specificity of 92.6\%

We also tried multi-class classification, which we predicted overdose vs. not overdose, as well as the drug that would be overdosed on, but our prediction errors were still quite poor and our models require further tuning.

\begin{figure}[h]
\centering
\includegraphics[width=0.75\textwidth]{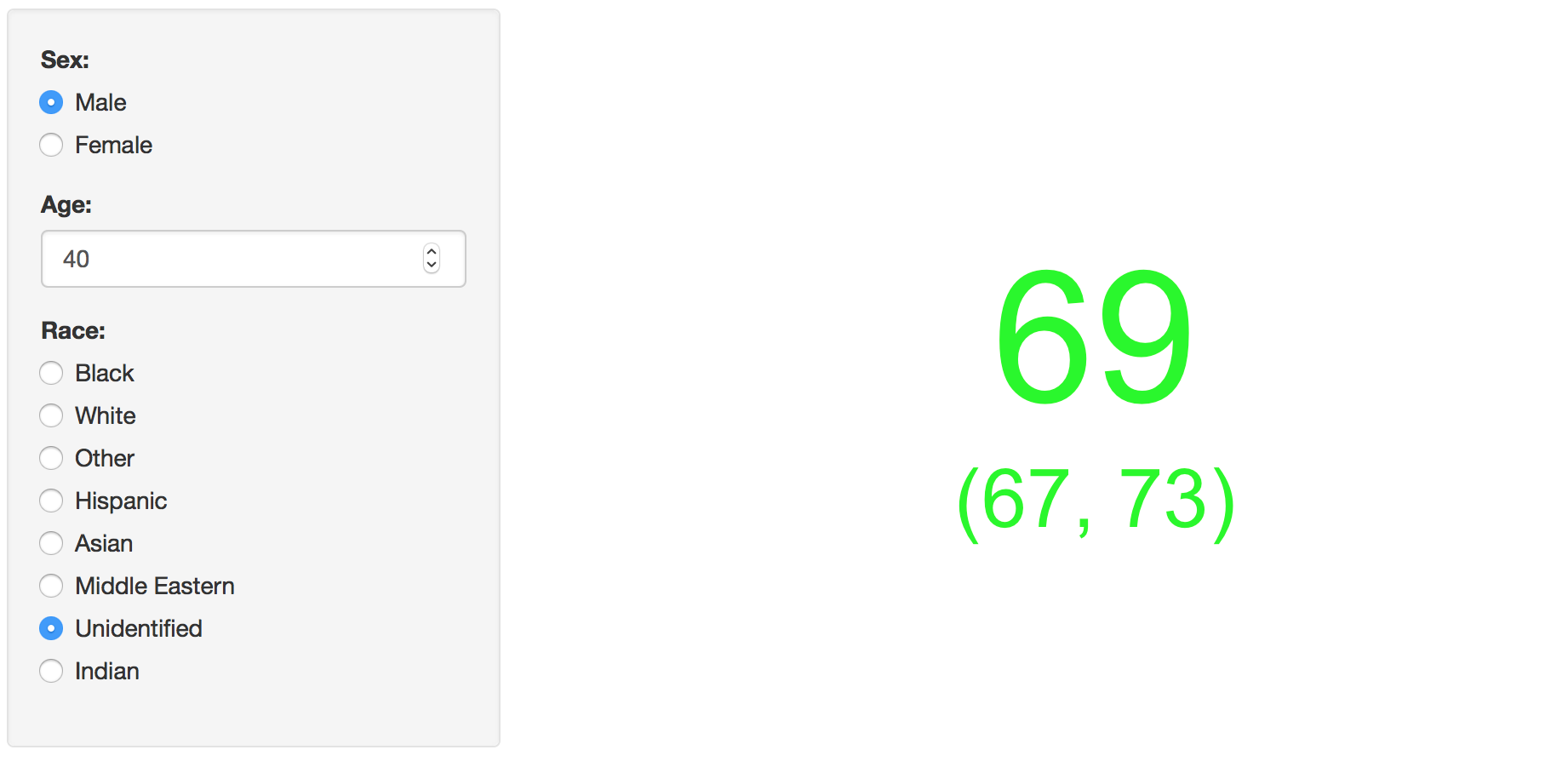}
\caption{Example of an interactive prediction tool. Given the set of covariates of a patient, it recalculates the risk score of overdose with a 95\% confidence interval.}
\label{fig:predictor}
\end{figure}

In particular, we applied random forest and neural network prediction algorithms to the dataset. We chose the neural network due to its known track-record of highly accurate predictions. We chose the random forest since it is also a efficient, effective algorithm, with the added benefit that it has more easily interpretable feature importance. For example, one can build the average decision tree from the random forest and show this to policy-makers. In our interactive prediction tool, we show how the relevant features can be used to calculate a score of risk overdose with a 95\% confidence interval ~\ref{fig:predictor}.

\section{Conclusion}
In this section, we use the results from our interactive data explorer and synthetic prediction tool to recommend policy changes. We also describe the major flaws in our current approach and future work to improve our models and analysis.

\subsection{Policy Recommendations}
Given the trends we discovered using our interactive data explorer, the characteristics of overdoses themselves and the individuals who did so, as well as the substantial improvement in our model after adding just five features minimally correlated with the outcome of interest, we have a few policy recommendations to slow and, ultimately, stop this opioid epidemic.
\begin{enumerate}
\item {\bf Collect more data and learn more about this problem.} First are foremost, more data is necessary in order to make more effective policy recommendations and assess individual risk factors. While we have preliminary evidence to suggest age and certain races are positively associated with the number of drugs used, our synthetic data analysis shows how powerful additional medical information could be in improving these methods.
\item {\bf Crack down on Fentanyl.} Again, more data is required to understand the more nuanced relationship here, but it appears increase in Fentanyl usage is a major driver of increased numbers of overdose deaths.
\item {\bf Combinations are killers.} Most overdose deaths are related to combinations of drugs being used, usually 2 or 3, but up to 6 or 7. Further analysis is required to understand which combinations specifically may be more dangerous, and which health conditions may make certain combinations more risky. It is also unclear whether numbers of drugs identified in overdose autopsies is higher now than a decade ago due to improved technology and precision in data collection, or because people are truly taking more drugs now.
\end{enumerate}
We also discovered Allegheny County has implemented a publicly available data explorer: \href{https://tableau.alleghenycounty.us/t/PublicSite/views/OverdoseinAlleghenyCounty/Home?\%3Aembed=y&\%3AshowAppBanner=false&\%3AshowShareOptions=true&\%3Adisplay_count=no&\%3AshowVizHome=no#5}{Overdose Deaths in Allegheny County}.

\subsection{Future Work}
One major area of improvement is optimizing our matching algorithm in order to provide a more representative synthetic dataset and associate covariates with the overdose cases. In particular, we would like:
\begin{itemize}
\item matching on more exhaustive criteria, such as geographical location (zip code), specific diagnoses within the past hospital history, or the number of hospital visits;
\item more flexible ways of drawing samples that allows criteria to match closely;
\item statistically sound way of dealing with missing data or skewed samples in our synthetic distribution.
\end{itemize}

One major area of further investigation are the following questions, which we did not have time to address in this analysis.
\begin{itemize}
\item Does the significant effect of white and Middle Eastern race on the number of drugs used disappear once socioeconomic status is taken into account?
\item Does the significant effect of age on the number of drugs used disappear once health is taken into account?
\item What are the bigger killers: hard drugs combined with alcohol, or a dangerous variety of less potent drugs?
\item What is the elbow point of drug usage: is there a particular drug or drug combination that leads to a significant spike in drug usage and overdose risk?
\item How correlated do additional features need to be to the outcome of interest in order to yield model improvements?
\item What masking techniques are most effective at providing high prediction accuracy improvements at low cost to privacy?
\end{itemize}

The opioid epidemic in Allegheny County, and in the United States more broadly, is an important problem. Better data exploration tools can increase policy maker understanding, and synthetic predictive tools can showcase the usefulness of access to additional data in this space.

\section*{Acknowledgements}

We would like to thank the organizers of the 2018 HackAuton hackathon: Nick Gisolfi, Chirag Nagpal, and the rest of the volunteers at \href{https://www.autonlab.org/}{Auton Lab}. We would also like to thank the generous sponsors who made this possible: The Carnegie Mellon University \href{https://www.cs.cmu.edu/}{School of Computer Science}, \href{https://www.ri.cmu.edu/}{The Robotics Institute}, \href{https://www.ml.cmu.edu/}{The Machine Learning Department}, and \href{https://www.heinz.cmu.edu/}{Heinz College}; as well as our corporate sponsors: \href{http://enterprises.upmc.com/}{UPMC Enterprises}, \href{https://www.usa.philips.com/}{Philips}, and \href{https://rxthinking.com/}{RxThinking}.

\bibliographystyle{plainnat}
\nocite{*}
\bibliography{references}

\end{document}